\documentclass[12pt]{iopart}

\usepackage[utf8]{inputenc}
\expandafter\let\csname equation*\endcsname\relax
\expandafter\let\csname endequation*\endcsname\relax

\usepackage{iopams}  
\usepackage{graphicx}
\usepackage{amsmath}

\usepackage{bm}
\usepackage{graphicx}
\usepackage{color}
\usepackage{physics}
\usepackage{siunitx}
\usepackage{bbold}
\usepackage{hyperref}
\usepackage{comment}

\DeclareSIUnit\angstrom{\text {Å}}

\definecolor{Blue}{rgb}{0.3,0.3,0.9}
\definecolor{Red}{rgb}{0.9,0.3,0.3}
\definecolor{Green}{rgb}{0.3,0.6,0.3}
\definecolor{Black}{rgb}{0.0,0.0,0.0}

\newcommand{\ee}{\text{e}}

\newcommand{\ii}{\text{i}}

\newcommand{\up}{\uparrow}
\newcommand{\dw}{\downarrow}
\renewcommand{\vec}[1]{\mathbf{#1}}

\begin{document}


\title{Monolayer transition metal dichalcogenides under finite-pulse polarized radiation}

\author{Alejandro S. G\'omez $^{1,2}$, Yuriko Baba$^{1,2}$, Francisco Dom\'{\i}nguez-Adame$^2$, Rafael A. Molina$^3$}
\address{$^1$ Condensed Matter Physics Center (IFIMAC), Universidad Autónoma de Madrid, E-28049 Madrid, Spain}
\address{$^2$ GISC, Departamento de F\'{\i}sica de Materiales, Universidad Complutense, E-28040 Madrid, Spain}
\address{$^3$ Instituto de Estructura de la Materia IEM-CSIC, Serrano 123, E-28006 Madrid, Spain}
\ead{alejandros.gomez@estudiante.uam.es}
\ead{yuriko.baba@uam.es}

\begin{abstract}

Recent advances in time-resolved angle-resolved photoemission spectroscopy have enabled access to ultrafast electron states and their spin dynamics in solids. Atomically thin transition metal dichalcogenides are paradigmatic two-dimensional materials where electron momentum and spin degrees of freedom are coupled, being suitable candidates for time-resolved spectroscopy studies. In this work, we present a thorough study of the electron dynamics when these materials are subject to an intense finite-pulse driving radiation. We extend the scope of the conventional Floquet engineering and rely of the so-called $t-t^{\prime}$ formalism to deal with driving fields described with two distinct time scales, namely the envelope amplitude timescale and the time period of the external field. The interplay between the finite-pulse timescales and the intrinsic properties of the electrons gives rise to transient valley polarization and dynamical modifications of band structures, revealed by the time-dependent circular dichroism of the sample. 

\end{abstract}

%
\vspace{2pc}
\noindent{\it Keywords}: Transition metal dichalcogenides, Floquet physics, time-dependent dichroism, finite pulses

%
%
\section{Introduction}

Floquet engineering has emerged as a powerful tool for dynamically tailoring the properties of quantum materials by applying time-periodic perturbations~\cite{Aoki2014,Basov2017}. This approach has been widely explored for its ability to manipulate novel electronic and topological phases in systems ranging from ultracold atoms~\cite{Eckardt2017,Goldman2016,Dotti2024} to solid-state materials~\cite{Wang2013b, Mahmood2016, McIver2019}.  However, most theoretical treatments of Floquet phenomena rely on the assumption of perfectly periodic driving fields, which may not accurately reflect practical experimental settings as usually large amplitudes of the driving field are required and ultrashort laser pulses need to be used. In this context, time-resolved ARPES measurements have revealed the dynamics of Floquet-Bloch states at the surface of three-dimensional topological insulators~\cite{Wang2013b,Mahmood2016}. Remarkably, recent advances in the time resolution of ARPES measurements~\cite{Aeschlimann2021, Reimann2018, Ito2023, Reimann2023} make it possible to address the dynamics of topological states with subcycle precision~\cite{Ito2023}. Interestingly, high field intensities achievable in ARPES experiments enhance the energy exchange of the Bloch electrons with the driving field, thus dominating over the dephasing mechanisms due to scattering with defects and phonons~\cite{Aeschlimann2021}.

Two-dimensional materials are particularly well-suited for exploring the concepts of Floquet engineering, as the issue of light penetration depth is inherently negligible. Atomically thin transition metal dichalcogenides (TMDs), with their rich electronic structures, strong spin-orbit coupling, and valley-selective optical response, offer a particularly intriguing platform for studying applications of Floquet engineering~\cite{Oka2019,Bao2022,Cao2024}. 
Experimental advances in Floquet engineering of 2D materials have demonstrated remarkable control over electronic and optical properties through periodic driving. In graphene, irradiation with circularly polarized light has been shown to induce a topological bandgap, effectively creating a Floquet-Chern insulator ~\cite{McIver2019} and demonstrating experimentally the concepts introduced in the seminal paper by Oka and Aoki ~\cite{Oka2009}.
Valley-selective excitations and coherent control of excitons have also been achieved in TMDs using tailored optical pulses, highlighting the potential for valleytronics applications ~\cite{Sie2015,Kobayashi2023}. Despite this success, it can be argued that the advances have been slow and difficult. Challenges such as heating and decoherence remain significant obstacles. Additional theoretical efforts are required to address more realistic experimental conditions, such as the finite duration of laser pulses and their impact on Floquet states.

In this work, we extend the scope of Floquet engineering applied to TMDs by considering finite-pulse driving schemes within the $t-t^{\prime}$ formalism~\cite{Peskin1993,Grifoni1998,Drese1999, Holthaus2015,Ikeda2022,Baba2024}. This formalism offers a natural framework to describe the temporal evolution of systems subjected to driving fields with arbitrary time profiles. It requires consideration of two distinct time scales: the envelope amplitude timescale and the time period of the external field. For the method to be effective, the period associated with the field frequency must be significantly shorter than the envelope timescale. In this way, the pulse driving is described using an instantaneous Floquet basis, with the pulse amplitude as a decoupled parameter. By applying this methodology to TMDs, we aim to elucidate how finite-pulse radiation influences the emergence of Floquet states, their spectral features, and the resulting valley dynamics. 

Our analysis reveals that the interplay between the finite-pulse characteristics and the intrinsic properties of TMDs gives rise to distinctive effects, including transient valley polarization and dynamical modifications of band structures that depend on the pulse shape and duration. We focus on the time-dependent circular dichroism as the experimental signature to detect these dynamical effects. These findings broaden the theoretical understanding of Floquet engineering applied to two-dimensional materials and offer experimentally relevant insights for designing optoelectronic devices and exploring ultra-fast valleytronics applications in TMDs.

\section{Method}

\begin{figure}[htb]
    \centering
    \includegraphics[width=\linewidth]{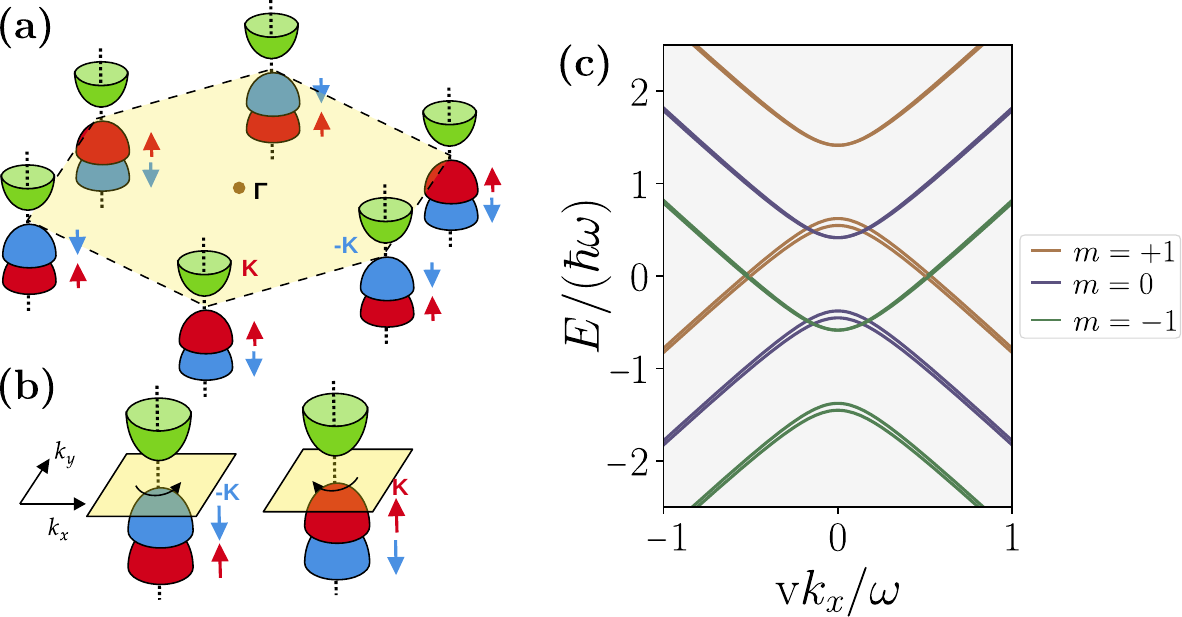}
    \caption{
    (a)~Schematic representation of the MoS\(_2\) band structure in the two-dimensional Brillouin zone close to the band edges. 
    Valley’s conduction bands (in green) are degenerated while valence bands (in red and blue) are split by the spin-orbit coupling (up/down arrows). 
    (b)~Enlarged view of the inequivalent valleys \(\tau = -1\) (left) 
    and \(\tau = +1\) (right), highlighting their opposite chiralities. 
    (c)~Calculated band spectrum of MoS\(_2\) including Floquet replicas $m=0,+1,-1$. The driving frequency is set to $\omega= 120$ THz, corresponding to a photon energy $\hbar \omega= 2\,$eV.
}
    \label{fig:Fig1Bands}
\end{figure}

In monolayer TMDs such as MoS$_2$, MoSe$_2$, WS$_2$ and WSe$_2$ the conduction- and valence-band edges  are located at valleys $\mathbf{K}$ and $\mathbf{K}^{\prime}$ of the hexagonal Brillouin zone~\cite{Lebegue2009,Zhu2011}. The conduction band is spin-degenerated, while the valence band shows a large spin splitting due to the enhanced spin-orbit coupling, see a schematic plot in figure~\ref{fig:Fig1Bands}(a). By comparison with \emph{ab-initio} calculations, it is shown that the electron states close to the band edges can be accurately described by the following two-band spinfull ${\bm k}\cdot {\bm p}$ Hamiltonian~\cite{Xiao2012}
\begin{equation}
  \mathcal{H}_0(k_x,k_y)=\hbar \mathrm{v} \left(\tau k_x \hat{\sigma}_x+k_y \hat{\sigma}_y\right)+\frac{\Delta}{2}\,\hat{\sigma}_z-\lambda \tau \, \frac{\hat{\sigma}_z-\hat{\sigma}_0}{2} \hat{s}_z~,
  \label{Hamiltoniano0}
\end{equation}
where $\tau=\pm 1$ is the valley index. Here $\hat{\sigma}_j$ ($j=x,y,z$) and $\hat{\sigma}_0$ denote the Pauli matrices and the $2\times 2$ unit matrix acting upon the basis functions. Similarly, $s_z$ and $s_0$ are the Pauli matrix and the $2\times 2$ unit matrix for spin. In equation~\eqref{Hamiltoniano0}, $\mathrm{v}$ is a model parameter with dimensions of velocity, $\Delta$ is the energy gap, and $\lambda$ is the spin splitting at the valence-band edge. 
The dispersion relation is: 
\begin{gather}
    E_{c \up (c\dw) }=\mathcal{M}_\tau \pm \frac{\lambda \tau}{2}~,   \quad 
     E_{v\up (v\dw)}=-\mathcal{M}_\tau\pm\frac{\lambda \tau}{2} ~,\\ 
    \mathcal{M}_\tau(k_x,k_y)=
    {\sqrt{(\hbar \mathrm{v})^2(k_x^2+k_y^2)+\frac{1}{4} \left(\Delta \pm \lambda \tau \right)^2}} ~, \nonumber
\end{gather}
where the subscript $c$ $(v)$ refers to the conduction (valence) band and the $\up~(\dw)$ to the spin up (down), which relates to the $+(-)$ sign. 

The external pulse is included via the Peierls substitution ${ \mathbf{k}} \to { \mathbf{k}} +\ee {\bm A}(t)$. 
The time-dependent Hamiltonian obtained is
\begin{subequations}
\begin{align} 
\mathcal{H}&(k_x, k_y, t)  =\mathcal{H}_0(k_x,k_y)
+ V(t)~, \label{eq:H(t)} \\
V&(t)  = \ee \hbar \mathrm{v} A(t) \left[\tau \cos(\omega t + \theta)\sigma_x+\sin(\omega t)\sigma_y \right]s_0~,
\label{eq:V(t)}
\end{align}
\end{subequations}
where $\theta$ gives the polarization of the light, being $\theta= 0$ for right-handed polarization and $\theta=\pi$ for left-handed, $\omega$ is the frequency of the pulse and the amplitude is given by a time-dependent function $A(t)$. 
Note that, due to the sign exchange between valleys the trajectory of ${\bm k} + e {\bm A}(t)$ around a
Dirac point in a circularly polarized light field is opposite for the two valleys,
see figure~\ref{fig:Fig1Bands}(b). 
For the sake of concreteness, we will employ a Gaussian-shape envelope, although any other differentiable function can be implemented within the method explained hereafter. Therefore
\begin{equation} \label{eq:Gaussian}
    A(t) = A_0 e^{-(t/\gamma)^2}~,
\end{equation}
where $A_0$ is the maximum amplitude and $\gamma>0$ has units of time and gives the width of the Gaussian pulse. 

\par Even when the pulse is not strictly periodic, we can define a Floquet-Fourier expansion at a fixed amplitude by considering two time-scales and writing equation~\eqref{eq:V(t)} as:
\begin{equation} \label{eq:V(t,tp)}
    V(t, t^{\prime})  =  \ee \hbar \mathrm{v} A(t^{\prime}) \left(\tau \cos(\omega t + \theta)\sigma_x+\sin(\omega t)\sigma_y \right)~.
\end{equation}
Notice that now the time-periodicity is restored for time $t$ leading to $V(t, t^{\prime}) = V(t+T, t^{\prime})$, where $T= 2\pi/\omega$ is the period of the internal oscillations of the pulse. The method of decoupling the envelope time-dependence from the frequency oscillation of the pulse is known as the $t-t^{\prime}$ formalism, and it is employed to describe the time-evolution of the system in a Floquet-like basis for non-periodic pulses~\cite{Peskin1993, Drese1999, Holthaus2015, Ikeda2022, Baba2024}. 
For a fixed $A(t^{\prime})$, the expansion over the Fourier harmonics is defined in the usual way
\begin{equation} \label{eq:H_FF}
    H_{mn} (t^{\prime})=
    \frac{1}{T} \int_T d t 
    \left[ \mathcal{H}_0(k_x,k_y) + V(t, t^{\prime}) 
    \right]
    e^{i(m-n)\omega t}
    - m \hbar \omega \delta_{m,n}~,
\end{equation}
where $m, n$ are integers corresponding to the \textit{harmonic} indices of the Fourier expansion. 
The Floquet-Fourier Hamiltonian is then obtained by writing the former equation in a matrix form as
\begin{equation} \label{eq:HF}
    H_F(\vec{k},A(t^{\prime})) =
\begin{pmatrix}
\ddots  &   Q(A(t^{\prime}))   & 0      &        &        \\
Q^\dagger(A(t^{\prime}))  & H_0(\vec{k})+\hbar \omega & Q(A(t^{\prime})) & 0 &         \\
0  &  Q^\dagger(A(t^{\prime})) & H_0(\vec{k}) & Q(A(t^{\prime})) & 0 \\
   &   0 & Q^\dagger(A(t^{\prime})) & H_0(\vec{k}) - \hbar \omega & Q(A(t^{\prime}))  \\
  &        & 0 & Q^\dagger(A(t^{\prime})) & \ddots  \\
\end{pmatrix}~,
\end{equation}
where the term that couples the Fourier replicas is the one of a monochromatic field and it is given as
\begin{equation} \label{eq:Qop}
    Q(A(t^{\prime})) = \ii\,\frac{ \ee \hbar \mathrm{v} }{2}\,{A}(t^{\prime})
    \left(
     \tau \ee^{-\ii \theta} \sigma_x + \sigma_y
    \right) s_0 ~.
\end{equation}
The solution of the eigenvalue problem for $H_F$~\eqref{eq:HF} for a fixed $A(t^{\prime})$ gives the Floquet quasi-energies $\xi^{(m)}_b(\vec{k},A)$ and the Fourier modes $\ket{u^{(m)}_b(\vec{k}, A)}$, with $b= \{ c_\up, c_\dw, v_\up, v_\dw \}$ denoting the four bands of the model and $m$ the Fourier component. 
The quasi-energies and the Floquet-Fourier modes define the so-called instantaneous Floquet basis
\begin{equation} \label{eq:FQttp:phiF}
   \ket{\phi_b^F(\vec{k},t)}= e^{-i\xi_b (\vec{k},A) t/\hbar}~\ket{u_b(\vec{k},A, t)}~,
\end{equation}
where the amplitude $A= A(t^{\prime})$, the quasienergy is fixed by $\lim_{A \to 0}  \xi_b({\vec{k}}, A) \to E_b ({\vec{k}})~,$ and
\begin{equation}\label{eq:FourierDecomp}
    \ket{u_b(\vec{k},A, t)} = \sum_{m= -\infty}^\infty e^{-im\omega t} ~\ket{u_b^{(m)}(\vec{k},A)}~.
\end{equation}
The Fourier components $u_b^{(m)}$ in \eqref{eq:FourierDecomp} can be used to define the time-averaged spectral function $\overline{\rho}(\vec{k}, E)$. For a fixed amplitude $A$, i.e. for a perfectly periodic Floquet system, this quantity is given by
\begin{equation}\label{spectral}
    \overline{\rho}(\vec{k}, E) = \sum_{\alpha} \braket{u_b ^{(m)}(\vec{k})}{u_b ^{(m)}(\vec{k})} \delta (\xi_b + m\hbar \omega - E)~.
\end{equation}
The time-averaged spectral function for MoS$_2$ is represented in figure~\ref{fig:timeavDOS}(a) in the limit of a small amplitude pulse. 
For later use, it is convenient to define the double index $\alpha \equiv (b,l)$ that labels the band $b$ and the replica $l$, and the following quantities
\begin{subequations}
\begin{align}
\xi_{\alpha}(\vec{k}, A) & \equiv  \xi_b(\vec{k}, A) + l \hbar \omega~,  \label{eq:FQttp:xi_bl} \\
\ket{u_{\alpha}(\vec{k},A,t)} & \equiv e^{i l \omega t} \ket{u_b(\vec{k},A,t)}~. \label{eq:FQttp:ut_bl}
\end{align}
\end{subequations}
An example of the replicas appearing in the Floquet spectra is shown in panel~\ref{fig:Fig1Bands}(c), where the two replicas with $m = \pm 1$ are represented for a fixed amplitude.

\par The $t-t^{\prime}$ formalism consists in mapping the time-evolution of the time-dependent Schrödinger equation (TDSE) for the full Hamiltonian~\eqref{eq:H(t)} given by 
\begin{equation} \label{eq:TDSE}
    i \hbar \frac{d}{dt}\,\ket{\psi(\vec{k},t)} = \mathcal{H}(\vec{k},t)\,\ket{\psi(\vec{k},t)}~,
\end{equation}
to the instantaneous Floquet basis by writing the solution as:
\begin{equation} \label{eq:FQttp:psit_dec}
   \ket{\psi(\vec{k},t)} = \sum_{\alpha} C_{\alpha} (\vec{k},t)\,\ket{u_{\alpha}(\vec{k}, A(t), t)}~,
\end{equation}
where the coefficients $C_\alpha$ satisfy
\begin{equation} \label{eq:FQttp:c_t}
    i \hbar \,\frac{d C_{\alpha}(\vec{k},t)}{dt} = \sum_{\beta} \mathcal{H}^{tt^\prime}_{\alpha \beta}(\vec{k}, A(t))\,C_\beta(\vec{k}, t)~,
\end{equation}
with
\begin{subequations}
\begin{align}
    \mathcal{H}^{tt^\prime}_{\alpha \beta} (\vec{k}, A(t)) & \equiv \delta_{\alpha \beta} \xi_{\alpha}(\vec{k}, A(t))
    - i \hbar\, \frac{dA}{dt} \,\mathcal{G}^{tt^\prime}_{\alpha \beta} (\vec{k},A(t))~,
    \label{eq:FQttp:Cham}\\
    \mathcal{G}^{tt^\prime}_{\alpha \beta} (\vec{k},A(t)) & \equiv  \frac{1}{T}\int_{T} dt^{\prime} \braket{u_\alpha(\vec{k},A(t),t^{\prime})|
    \partial_A}{u_\beta(\vec{k},A(t),t^{\prime})}~.
    \label{eq:FQttp:Gab}
\end{align}
\end{subequations}
Here, $\partial_A = \partial/\partial A$. 
The instantaneous Floquet states in \eref{eq:FQttp:Gab} are assumed to be differentiable, and we chose the gauge of the parallel transport over the states by setting $\braket{u_\alpha(A,t)|\partial_A}{u_\alpha(A,t)}=0~$. More details about the implementation of the $t-t^{\prime}$ formalism can be found in reference~\cite{Baba2024}.

\par Finally, to relate the calculated coefficients to an observable quantity, we define the integrated power over time of the $C_\alpha(\mathbf{k},t)$ coefficients as

\begin{equation}\label{P(Omega)}
    P(\Omega) = \int d \mathbf{k}\int dt \sum_\alpha |C_\alpha(\vec{k},t)|^2 L(\Omega,\varepsilon_\alpha, \Gamma)~,
\end{equation}
where we are integrating in a time-window such that the amplitude $A > A_\mathrm{min}$. In the previous expression $L(z, z_0, \Gamma)$ is the Lorentzian function defined as:
\begin{equation}
    L(z, z_0, \Gamma) = \frac{1}{\pi}\,\frac{\Gamma}{(z - z_0)^2 + \Gamma^2}\; .
\end{equation}
%
As a reference of the initial state, we define the limit of $P(\Omega)$ integrated for $A_0\to 0$ or, equivalently, upon substitution of the time integral at $t\to t_0$. In this limit, we obtain
\begin{align}
P_0(\Omega) & = \int d\mathbf{k} \, \int d t\, \sum_\alpha \abs{C_\alpha(\mathbf{k},t_0)}^2 L(\Omega,\varepsilon_\alpha, \Gamma) \delta(t-t_0) \nonumber \\ 
& = \sum_\alpha \int d\mathbf{k}\,   
|\braket{u_\alpha(\vec{k})}{\psi (t= t_0)}|^2 L(\Omega,\varepsilon_\alpha, \Gamma)~,        
\end{align}
The quantity $P_0(\Omega)$ concerns the initial distribution of replica weights given an initial state $\psi_0$ and it can be employed as a comparison with $P(\Omega)$, which accounts for the weights after the pulse has been applied to the system.

Finally, to quantify the effect of the pulse polarization, we analyse the circular dichroism of this integrated power~\cite{cao2012valley,yao2008valley}:
\begin{equation}\label{eq:CD}
    \mathrm{CD} (\Omega)= \frac{ \abs{P^\circlearrowright}-\abs{P^\circlearrowleft}}{\abs{P^\circlearrowright}+\abs{P^\circlearrowleft}}~,
\end{equation}
where the superscript stands for light polarization, being $\theta = 0$ the right-handed ($\circlearrowright$)and $\theta = \pi$ the left-handed ($\circlearrowleft$) light polarization.


In this work, and for all subsequent numerical simulations, we consider molybdenum disulphide (MoS\(_2\)) as the material platform. The adopted parameters are: Fermi velocity $ \hbar\mathrm{v} = 3.512\, \mathrm{eV}$\AA$^{-1}$, bandgap \( \Delta = 1.66\,\mathrm{eV} \), and spin-orbit coupling \( \lambda = 0.15\,\mathrm{eV} \) \cite{liu2015electronic}. The driving frequency is set to \( \omega = 240~\mathrm{THz} \), corresponding to a photon energy \( \hbar\omega = 1\,\mathrm{eV} \). If not stated otherwise for the integrated power $P(\Omega)$ the cut-off amplitude employed is $A_\mathrm{min}/(\hbar\omega) = 0.01$.

\section{Results and discussion}


\par
\begin{figure}[hbt]
    \centering
    \includegraphics[width=0.8\textwidth]{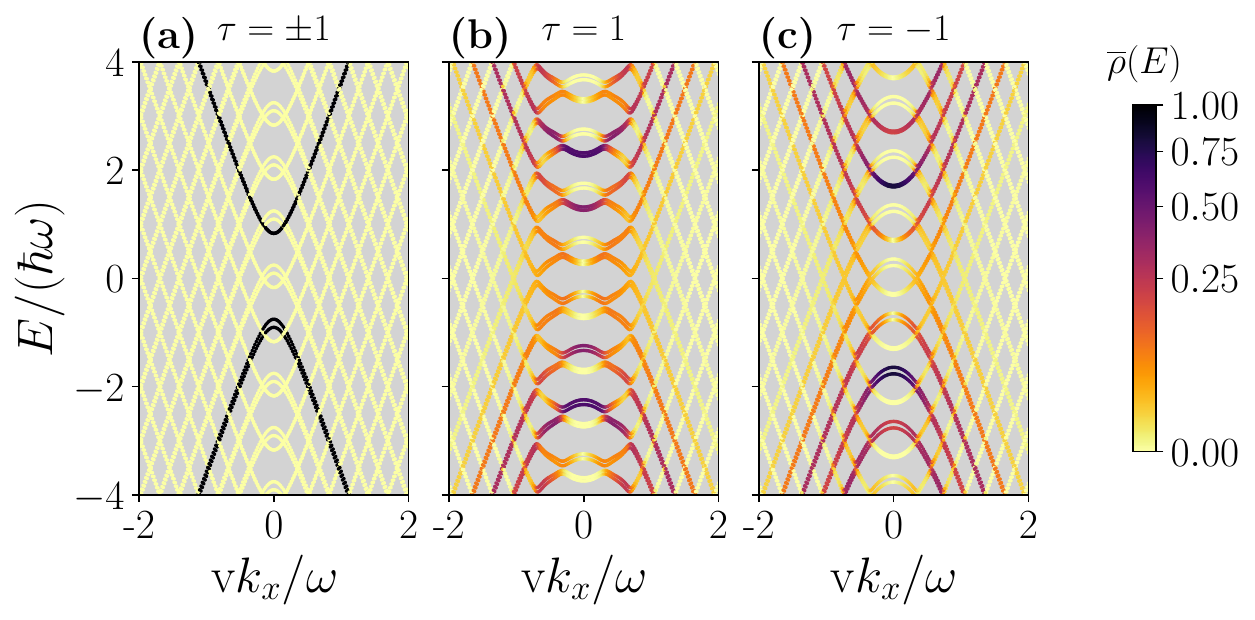}
    \caption{
    Floquet spectrum for a fixed pulse amplitude $A_x=A_y=0$ in panel (a) and $ A_x = A_y = 0.5 \hbar \omega$ in panels (b) and (c). The valleys are indicated on top of the figure, both inequivalent valleys $\tau = \pm 1$  are considered. The colour code corresponds to the time averaged DOS \(\overline{\rho}(E)\) computed according to~\eqref{spectral}. 
    The pulse polarization is set to a right-handed circular polarization, i.e. setting $\theta = 0$ in \eqref{eq:V(t)}. The spectra are plotted for $k_y=0$.
    }
    \label{fig:timeavDOS}
\end{figure}

 First, we briefly analyse the effect of the circularly polarized pulse on the valleys in the case of a constant amplitude. 
To show the distinct behaviour of $\tau=\pm 1$ valleys, we fix the external pulse to a right-handed circularly polarized monochromatic field, i.e. setting $\theta = 0$ in \eqref{eq:V(t)}. The results for MoS\(_2\) are shown in figure~\ref{fig:timeavDOS}, where the time-averaged density of states \( \overline{\rho}(E) \) from~\eqref{spectral} is computed for a fixed amplitude $A$. 
The dichroism is visible in the distinct spectral features between the \textbf{K} ($\tau=+1$) and \textbf{K'} ($\tau=-1$) valleys, as evidenced by the contrast in the bandgap magnitudes appearing between Floquet replicas in figure~\ref{fig:timeavDOS}(a)--(b). Specifically, the valley with chirality aligned to the pulse's handedness, in this case $\tau = +1$, exhibits significantly larger pulse-induced gaps. For the other valley, the spectral population is displaced to the replicas, but the Floquet quasienergies are not dramatically modified by the pulse, compare panel~\ref{fig:timeavDOS}(b) with the spectrum for $A=0$ in panel~\ref{fig:Fig1Bands}(c). However, even if the gaps in the $\tau=-1$ are almost negligible, it is important to stress that, due to the form of the time-dependent perturbation, $V(t)$ and the unperturbed Hamiltonian $\mathcal{H}_0$ do not commute and hence all band crossings between replica bands are avoided. 
The commutator is in fact given by
\begin{equation}\label{NoConmut}
\begin{aligned}
     \left[\mathcal{H}_0,V(t)\right] 
    =\ii \,\ee \hbar \mathrm{v}  A(t) \left\lbrace 
    2\hbar \mathrm{v} \tau \left[k_x \sin (\omega t)-k_y \cos(\omega t + \theta)\right] \hat{\sigma}_z \otimes \hat{s}_0 \right. \\ \left. + \left[
    \tau \cos(\omega t + \theta)\hat{\sigma}_y-\sin(\omega t)\hat{\sigma}_x
    \right] \otimes \left[\Delta \hat{s}_0-\lambda \tau \hat{s}_z\right]
    \right\rbrace \nonumber ~, &
\end{aligned}
\end{equation}
and hence it is generally non zero. 


\par
\begin{figure}[hbt]
    \centering
    \includegraphics[width=\linewidth]{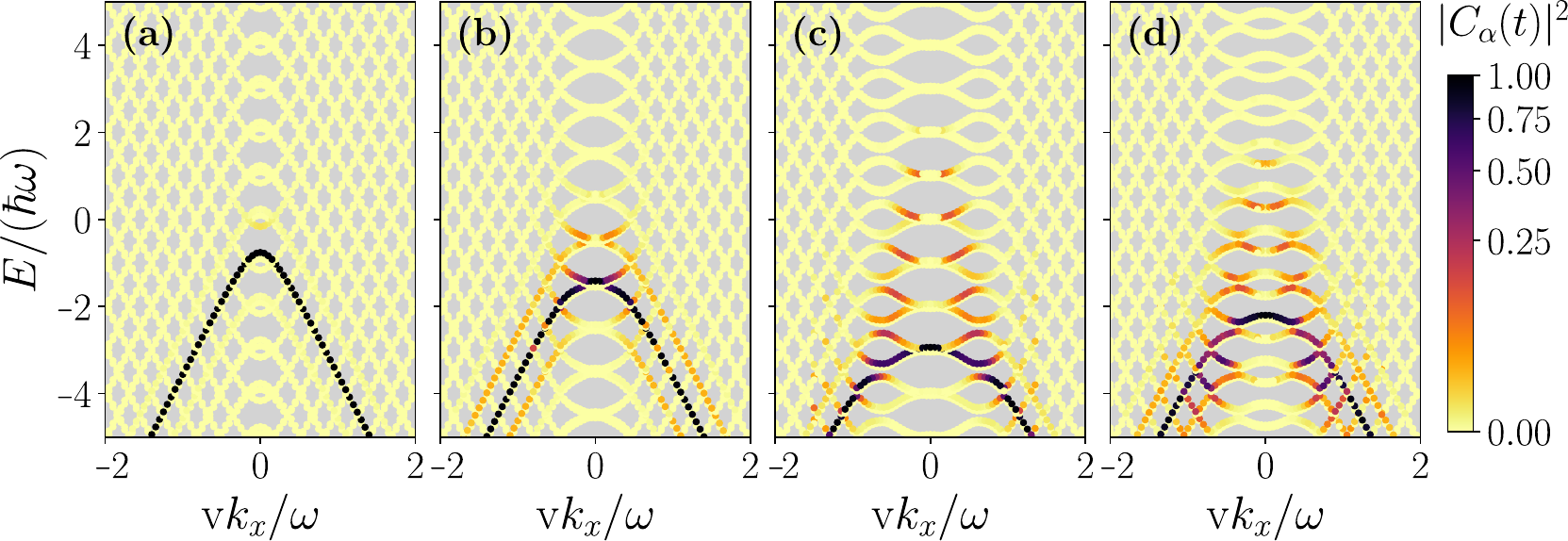}
    \caption{Time-dependent Floquet spectra for the valley $\tau=+1$ under a finite-pulse with right-handed circular polarization ($\theta=0$) and Gaussian envelope centred at $t_0=0$ with a width $\gamma=T$, where $T$ is the pulse period. 
    The spectra are shown for $t/T=\{-2,-1,0,0.6 \}$ in panels (a), (b), (c) and (d), respectively.
    We consider as initial state a fully occupied valence band with spin up $\ket{v_\uparrow^0}$
    and a maximum amplitude $A_0/(\hbar\omega) =0.7$.
    The colour code corresponds to $|C_\alpha(\mathbf{k},t)|^2$. 
    }
    \label{fig:bandsTTP}
\end{figure}

\begin{figure}[h!]
    \centering
    \includegraphics[width=\linewidth]{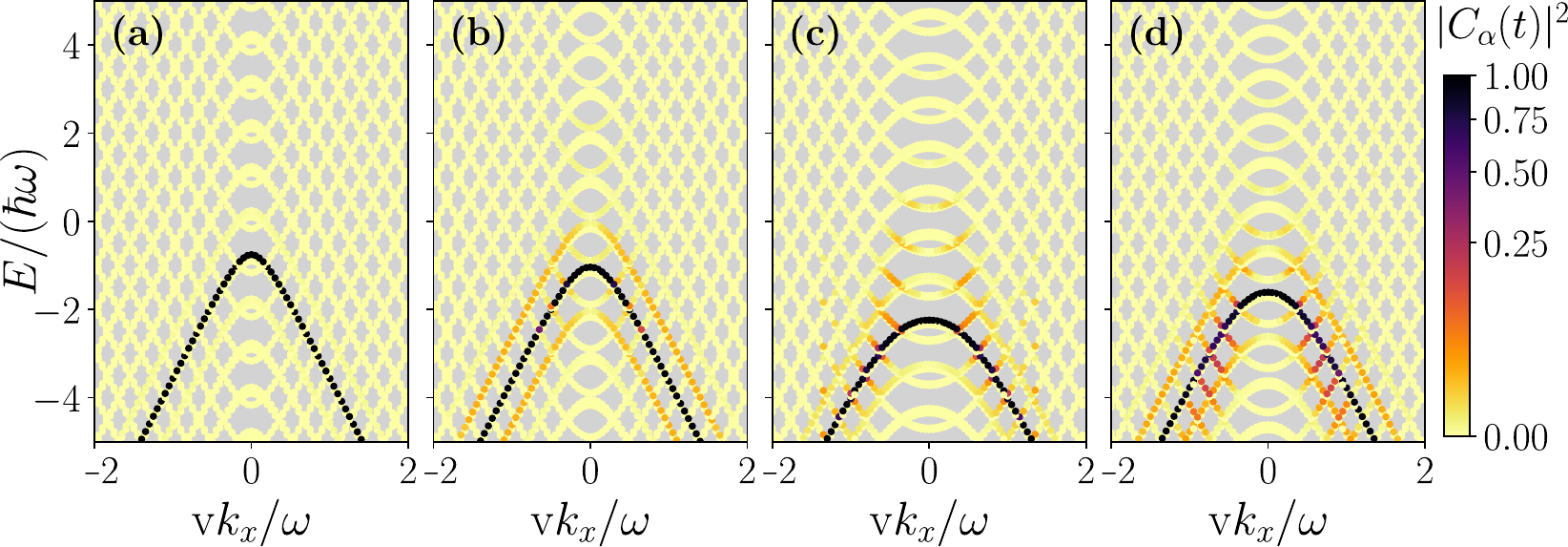}
    \caption{Same magnitudes as in figure~\ref{fig:bandsTTP} for the valley $\tau=-1$ considering as initial state $\ket{\psi_0}=\ket{v_\downarrow^0}$. The rest of the parameters are the same as in figure~\ref{fig:bandsTTP}. }
    \label{fig:bandsTTP_tauneg}
\end{figure}

 Next, we consider a finite pulse with Gaussian envelope given by~\eqref{eq:Gaussian}. Within the 
$t-t^\prime$ formalism, the time-dependent Floquet spectra are computed in figures~\ref{fig:bandsTTP} and \ref{fig:bandsTTP_tauneg} for the two valleys~$\tau=\pm 1$ and fixed right-handed circular polarization, i.e. $\theta = 0$ in \eqref{eq:V(t)}. 
The time-dependent Floquet spectrum at time $t$ resembles the corresponding time-periodic Floquet spectrum for constant amplitude $A(t)$. Therefore, the gapped structure between replicas is indeed enhanced only for the valley $\tau=+1$ in figure~\ref{fig:bandsTTP}. For $\tau=-1$ the bands are displaced and renormalized by the pulse even if the gaps between replica bands are almost negligible, see figure~\ref{fig:bandsTTP_tauneg}. 
The figures include in colour code the expansion coefficients $C_\alpha(\mathbf{k},t)$ for some selected times.
Since $|C_\alpha(\mathbf{k},t)|^2$ is related to the occupation of the Floquet replicas, it can be seen that the valley in which the chirality matches the polarization of the pulse, in this case $\tau=+1$, the distributions of $C_\alpha(\mathbf{k},t)$ coefficients achieve higher replica bands. Thus, the effect of the pulse when polarized in the valley direction is enhanced not only at the level of the appearance of gaps in the spectrum, but also in terms of populating higher replica states. 
The additional figures~\ref{fig:spectrum_b1} and~\ref{fig:spectrum_b2} in the Appendix~\ref{sec:App} show additional time snapshots for completeness.

\par So far, we have discussed the impact of a fixed pulse on the two inequivalent valleys. However, a more convenient way of exploiting the effect of the external perturbation is to vary the external pulse itself so that the dynamics of the states can be tuned at will. 
In the case of the Gaussian pulse considered, the two parameters that can be modified are the Gaussian width $\gamma$ and the polarization angle $\theta$. 
From equations~\eqref{eq:V(t)} and~\eqref{Hamiltoniano0}, it can be seen that the Hamiltonian~\eqref{eq:H(t)} is invariant under the following transformation
\begin{align*}
    \tau \to - \tau~,\quad
    \theta \to \theta +\pi~,\quad
    k_x \to -k_x\quad
    s_z \to -s_z~.
\end{align*}
Thus, the exchange of the valley index is equivalent to inverting the pulse polarization, as well as the momentum $k_x$ and the spin of the state. 
For this reason, in the following, we discuss the influence of the pulse parameters for a fixed valley polarization $\tau = +1$, recalling that the results are equivalent for $\tau=-1$ upon the transformation above mentioned. 

\par
\begin{figure}[htb]
    \centering
    \includegraphics[width=\linewidth]{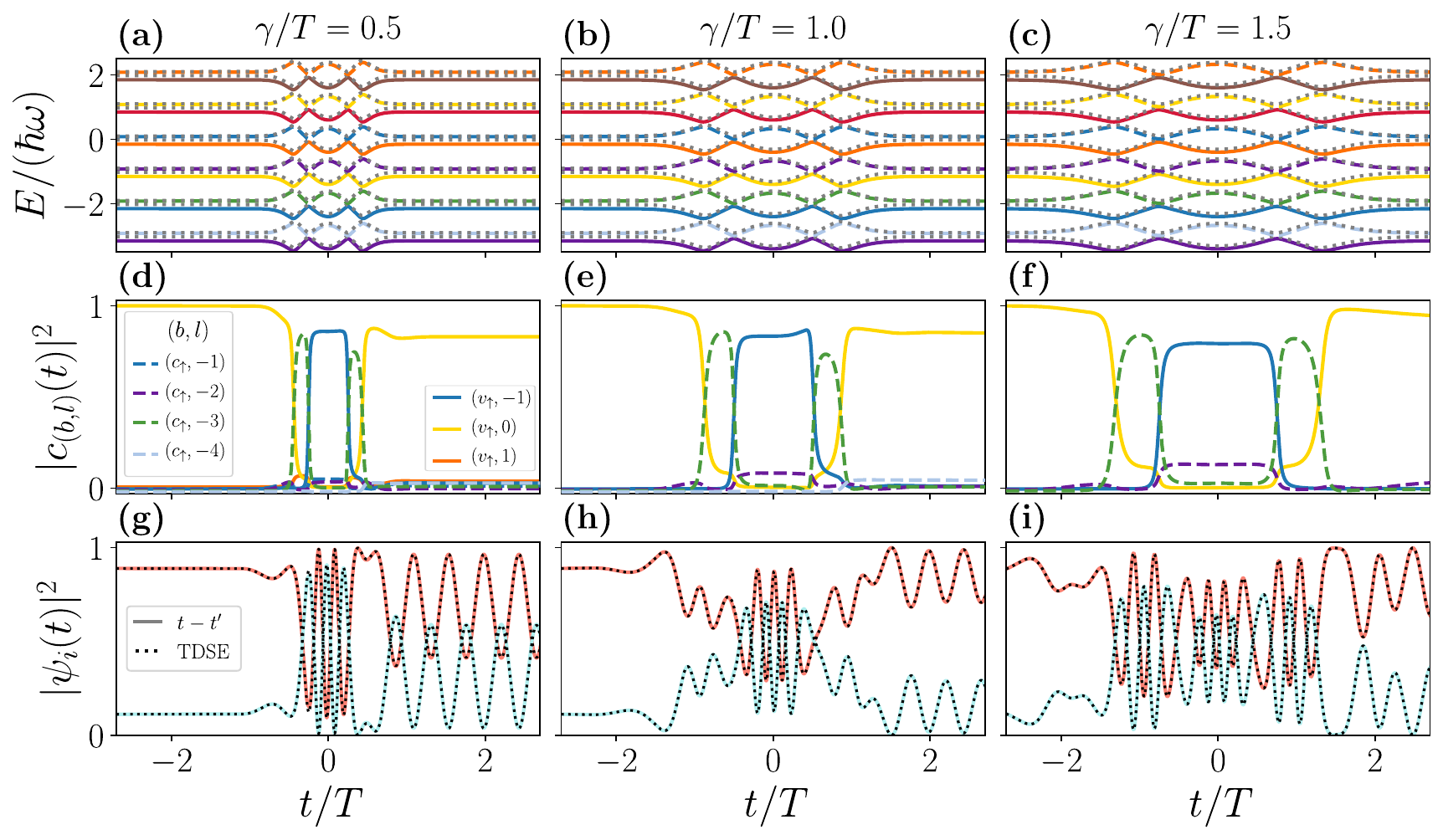}
    \caption{
    For \(\tau = +1\) and right-handed polarization we evaluate the time evolution of a single \(\mathrm{v}/\omega(k_x ,  k_y )=(0.2,0)\) mode, initialized in the state 
    \(\ket{v^0_\uparrow}\) of the unperturbed Hamiltonian. The width of the pulse 
    is varied as \(\gamma/T = \{0.5,\,1.0,\,1.5\}\). 
    Panels~(a), (b) and (c) show the Floquet spectra; 
    panels~(d), (e) and (f) display the squared probability 
    amplitudes of the coefficients \(c_{(b,l)}\); 
    and panels~(g), (h) and (i) compare the time evolution 
    derived from the \(t - t^{\prime}\) formalism to that obtained from the direct integration of TDSE, expressed in the orbital basis. 
    The maximum amplitude is set to \(A_0 = 0.5 \hbar \omega\). 
}
    \label{fig:gamma}
\end{figure}

 The evolution within the $t-t^{\prime}$ formalism of a single \((k_x, k_y)\) mode at fixed valley \(\tau = +1\) is shown in figure~\ref{fig:gamma} for different values of the width of the Gaussian envelope $\gamma$, corresponding to the columns of the figure. The initial state considered is \(\psi_0=\ket{v^0_\uparrow}\), corresponding to the spin-up valence state of the unperturbed Hamiltonian. 
The time-dependent Floquet quasi-energies are shown in panels~\ref{fig:gamma}{(a)--(c)}, the squared probability amplitudes of the expansion coefficients \(c_{(b,l)}\)  in panels~\ref{fig:gamma}{(d)--(f)}, and a comparison between the time evolution obtained via the $t - t^{\prime}$ formalism and the direct integration of the TDSE is plotted in~\ref{fig:gamma}{(g)--(i)}.
The widths of the pulse are $\gamma/T = \{ 0.5, 1.0, 1.5\}$. 
For each pulse width \(\gamma\), we observe the same qualitative evolution of the band populations, with the overall duration of these dynamics increasing as \(\gamma\) becomes larger, compare panels~\ref{fig:gamma}(d)--(f). The finite pulse does not mix spin states, and the distribution of Floquet replicas causes the system to alternate between conduction and valence states. Furthermore, as shown in figure~\ref{fig:gamma}, the width of the pulse also modifies the final state, while the specific Floquet replicas involved remain fixed by the maximum amplitude $A_0$ and the frequency $\omega$, which determine the structure of the Floquet couplings and are unchanged for the three cases considered. 

\par The comparison of the three pulse lengths in figure~\ref{fig:gamma} is a clear example of the 
advantage of employing the $t-t^{\prime}$ formalism to describe the dynamics of Hamiltonians with an underlying Floquet-like structure. In fact, by employing the Floquet picture, the system’s evolution can be understood as a superposition of states with energies \( E = \xi_b + m \hbar \omega\) which are populated according to the evolution of the Gaussian envelope.
By analysing panels~\ref{fig:gamma}(d)--(f) we can distinguish mainly two regimes in the evolution as a function of time: 
\begin{itemize}
    \item the adiabatic phase acquisition within  the same Floquet replica, which corresponds to the plateaus in the $|c_{(b,l)}|$ coefficients;
    \item the non-adiabatic transition between replicas, which correspond to the fast transitions between different $|c_{(b,l)}|$ coefficients. This second scenario is obtained in correspondence with avoided crossings of the time-dependent Floquet quasienergies, see panels~\ref{fig:gamma}(a)-(c). Formally, in the $t-t^{\prime}$ formalism, the non-adiabatic transitions are obtained when the term $\mathcal{G}^{tt^\prime}_{\alpha \beta}(\vec{k},A(t))$ leads the time-evolution in~\eqref{eq:FQttp:c_t}. 
\end{itemize} 
Hence, increasing the pulse amplitude broadens these plateaus, reflecting the extended time during which the system stays in a given Floquet-like state.
Finally, it is important to stress that the evolution in the $t-t^{\prime}$ formalism is indeed equivalent to the solution of the TDSE, as it can be seen by the comparison of the lower panels of figure~\ref{fig:gamma}. 
In the $\psi(t)$ expressed in the orbital bases, the interference generated by the pulse can be seen in  the oscillatory pattern as a function of $t$, which clearly shows distinct frequency regions. These distinct frequency regions correspond to the plateaus in panels~(d)--(f), so that the change in the oscillation pattern of $\psi(t)$ in panels~(g)--(f) can be explained by the time-dependent Floquet evolution of the expansion coefficients $c_{(b,l)}$.


%


\par
\begin{figure}[hbt]
    \centering
    \includegraphics[width=\linewidth]{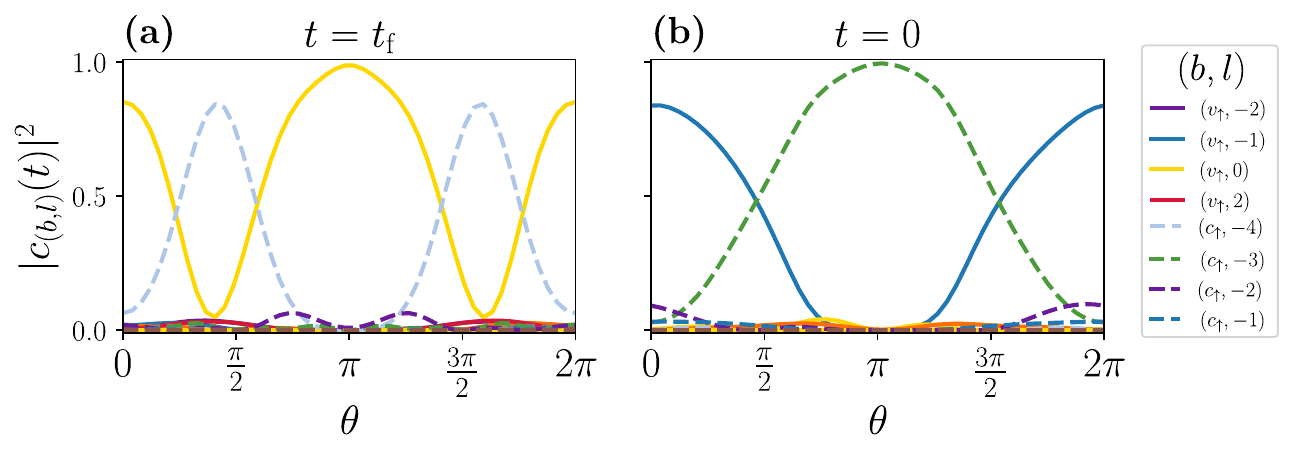}
    \caption{
    Squared amplitudes of the non-zero conduction and valence band coefficients 
    \(\lvert c_{(b,l)}(t) \rvert^2\) as a function of the polarization angle \(\theta\) 
    for a single mode \(\mathrm{v}/\omega(k_x ,  k_y )=(0.2,0)\), at \(\tau = +1\) valley and initialized state of
    \(\ket{v^0_\uparrow}\). 
    Panel~(a) shows the final-state amplitudes at \(t = t_f\). 
    Panel~(b) shows the amplitudes at the Gaussian peak time 
    \(t = 0\).
    The pulse width is fixed at \(\gamma = 1\) and maximum amplitude \(A_0/(\hbar \omega) = 0.5 \). 
}
    \label{fig:theta}
\end{figure}

 In figure~\ref{fig:theta} we analyse the dependence on the polarization angle. For the sake of concreteness, we show the results for the same \((k_x,k_y)\) mode described earlier, this time fixing the pulse width to \(\gamma = 1\) and varying the polarization angle \(\theta\). 
In panel~\ref{fig:theta}(a), we plot the final-state expansion coefficients \( \lvert c_{(b,l)}(t_f) \rvert^2 \) for a state initialized in $\ket{v^0_\uparrow}$ of the unperturbed Hamiltonian. 
In the limiting case of $\theta = \pi$ (left-handed polarization), the state returns to the initial state, consistently with the left-handed polarization not matching the chirality of the \(\tau = +1\) valley and hence not hybridizing the Floquet replicas.
In the case of the right-handed polarization, i.e. for $\theta = 0$, the state is driven back to the initial valence state with high, but not unit, probability. Part of the occupation is indeed displaced to the conduction replica $(c_\uparrow, -4)$. This result is consistent with the right-handed polarization matching the chirality of the \(\tau = +1\) valley and hence generating a stronger coupling between replicas and wider gaps between conduction and valence sidebands. 
Interestingly, as $\theta$ deviates from these limiting left and right-handed cases, the transition to this lower conduction-band replica (without spin flipping) is increased, demonstrating the capability to select particular final states simply by adjusting the circular polarization angle.

\par In figure~\ref{fig:theta}(b) we show the same squared amplitudes at the time of maximum pulse amplitude, \(t = 0 \), corresponding to the peak of the Gaussian envelope. Here, the effect of the light’s chirality on the final state becomes more pronounced. When the polarization angle \(\theta\) matches the valley chirality $\theta = 0$, the system is driven to an excited valence state. Conversely, for the opposite polarization, a conduction state is induced, once again illustrating the ability to control and select the evolution of the states and the transitions between sidebands via the polarization angle of the incident light.

\par
\begin{figure}[hbt]
    \centering
    \includegraphics[width=\linewidth]{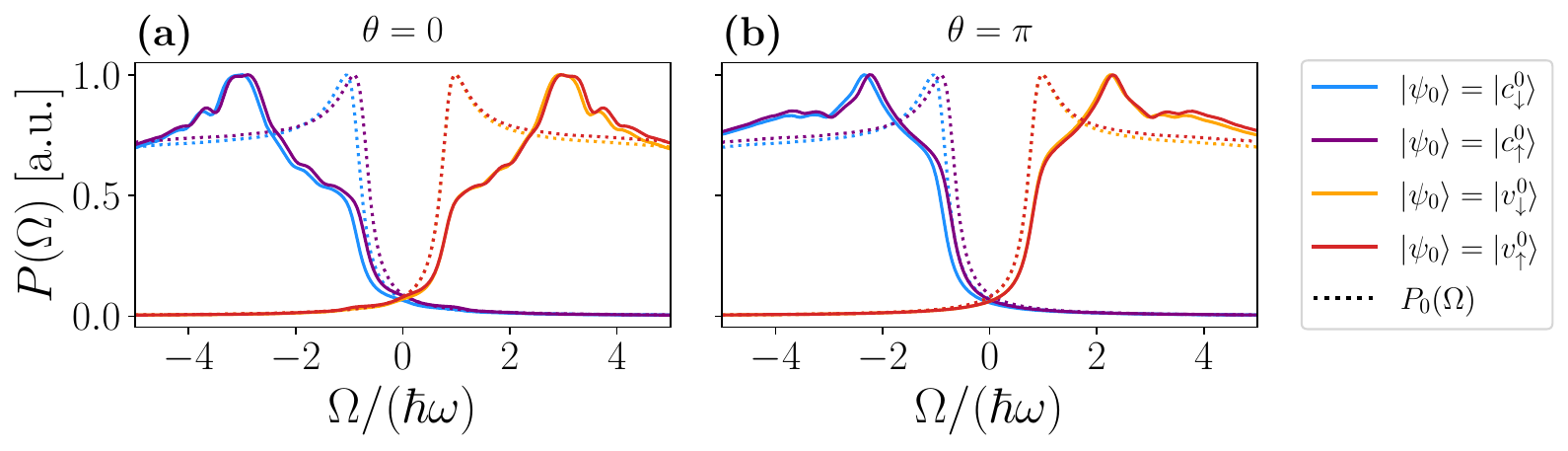}
    \caption{
    Integrated power \(P(\Omega)\) for a fixed valley \(\tau = +1\) 
    and pulse amplitude \(A_0/(\hbar\omega) = 0.5\,\). 
    The dotted line \(\,P_0(\Omega)\) indicates the initial distribution 
    of replicas before applying the pulse. 
    Panels~(a) and (b) compare different initial states 
    for right-handed (\(\theta=0\)) and left-handed (\(\theta=\pi\)) 
    circular polarization, respectively.
}

    \label{fig:Pomega}
\end{figure}

 To gain insight into the distribution for all $\vec{k}$-modes, we evaluate \(P(\Omega)\), defined in equation~\eqref{P(Omega)}, by integrating over all  $k_x$ momenta and times. We consider a fixed valley \(\tau = +1\) and a pulse amplitude \(A/(\hbar \omega) = 0.5 \). The dotted curve \(P_0(\Omega)\) indicates, as a reference, the initial distribution before the pulse is applied. In figure~\ref{fig:Pomega}(a), we present the integrated power for several initial states when the polarization of the pulse matches the chirality of the chosen valley. In addition, in figure~\ref{fig:Pomega}~(b), the polarization is opposite to the valley chirality. When the polarization and valley chirality coincide, figure~\ref{fig:Pomega}(a), \(P(\Omega)\) displays broader distributions in higher-energy (Floquet) replicas, indicating that higher sidebands are excited, along with a noticeable finer structure created by the wider gaps arising between sidebands. Conversely, for opposite chirality and polarization, in figure~\ref{fig:Pomega}~(b), the system does not populate higher replicas so effectively and the structure of the peaks is broader, indicating that the gaps between sidebands are almost zero.

\par
\begin{figure}[hbt]
    \centering
    \includegraphics[width=0.7\linewidth]{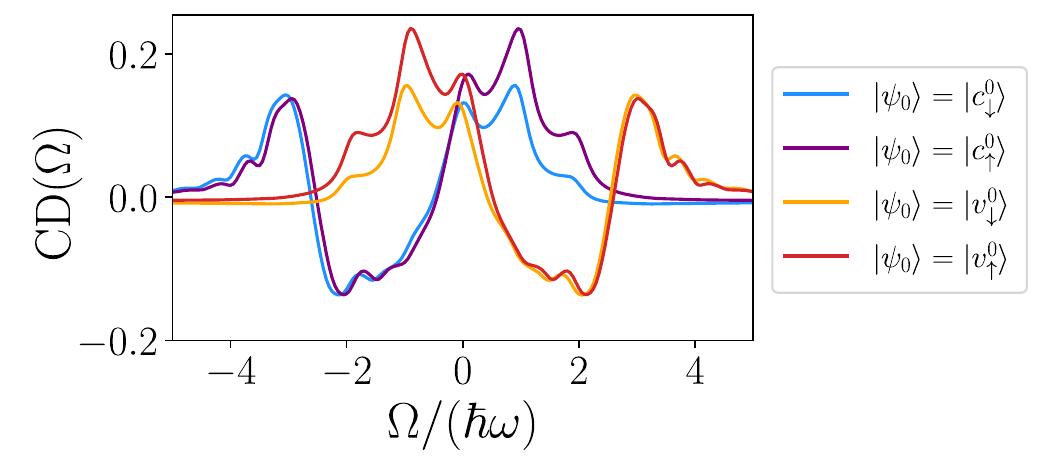}
    \caption{
    Circular dichroism \(\mathrm{CD}(\Omega)\) according to equation~\eqref{eq:CD} for $\tau=+1$ and $A_0/(\hbar\omega) = 0.5$. Each curve corresponds to a different initial state indicated in the legend.
}
    \label{fig:CD}
\end{figure}

 Finally, the circular dichroism (CD) is computed in figure~\ref{fig:CD} using~\eqref{eq:CD} for the valley \(\tau = +1\). Due to symmetry, an equivalent result holds for the other valley upon reversing the circular polarization direction. We observe several pronounced peaks in \(\mathrm{CD}(\Omega)\) 
that are a signal of the difference in the structure of the gaps between replicas and the population of (and transitions to) higher-energy states under right-handed versus left-handed circular polarization. 
The sign of the peaks indicates which polarization is more strongly coupled as a function of the energy, reflecting the selective excitation of states with matching valley chirality. 
Thus, the appearance and magnitude of these CD peaks highlight how polarization control can be leveraged to manipulate valley-selective processes and tune the occupation of higher Floquet sidebands even if finite-short pulses are employed.

\section{Conclusions}

In this work we have studied the effect of short pulses of radiation on single-layer TMDs. In particular, we examine pulses that can be described by two well-separated time-scales, one corresponding to the envelope of the amplitude and another related to the fast oscillations, which we consider of a fixed frequency $\omega$. Employing the two time-scales, a time-dependent Floquet theory, called Floquet $t-t^{\prime}$, can be formulated such that the Floquet states for each time-dependent amplitude value are employed as a basis to solve the time evolution of a given initial state. 
Within this formalism, the dynamics of the states can be interpreted in a time-dependent Floquet expansion, such that the evolution can be divided into two main regimes: the adiabatic acquisition of a phase and the non-adiabatic transition between Floquet sidebands. 
For the pulse amplitudes considered here, we find that the Floquet $t-t^{\prime}$ formalism can be successfully employed even for pulses with few oscillations within the envelope function. 

\par Regarding the results for the TMDs under circularly polarized irradiation, we have shown that the polarization angle $\theta$ can be employed to manipulate valley-selective processes related to the occupation of Floquet sidebands. In fact, if the polarization matches the valley handedness, the Floquet sidebands are strongly coupled and quantities such as the integrated power $P(\Omega)$ or the circular dichroism $\mathrm{CD}(\Omega)$ show a structured peak shape as a function of the energy, corresponding to the induced gaps between sidebands, as well as the displacement of the population from the initial state to the higher Floquet replicas. In this way, the external pulse parameters, such as the pulse width or the polarization angle, can be employed to obtain valley-selective dynamics of the states and to probe the dichroic nature of these materials. 

\section*{Appendix} \label{sec:App}

\begin{figure}[htb]
    \centering
    \includegraphics[width=0.7\linewidth]{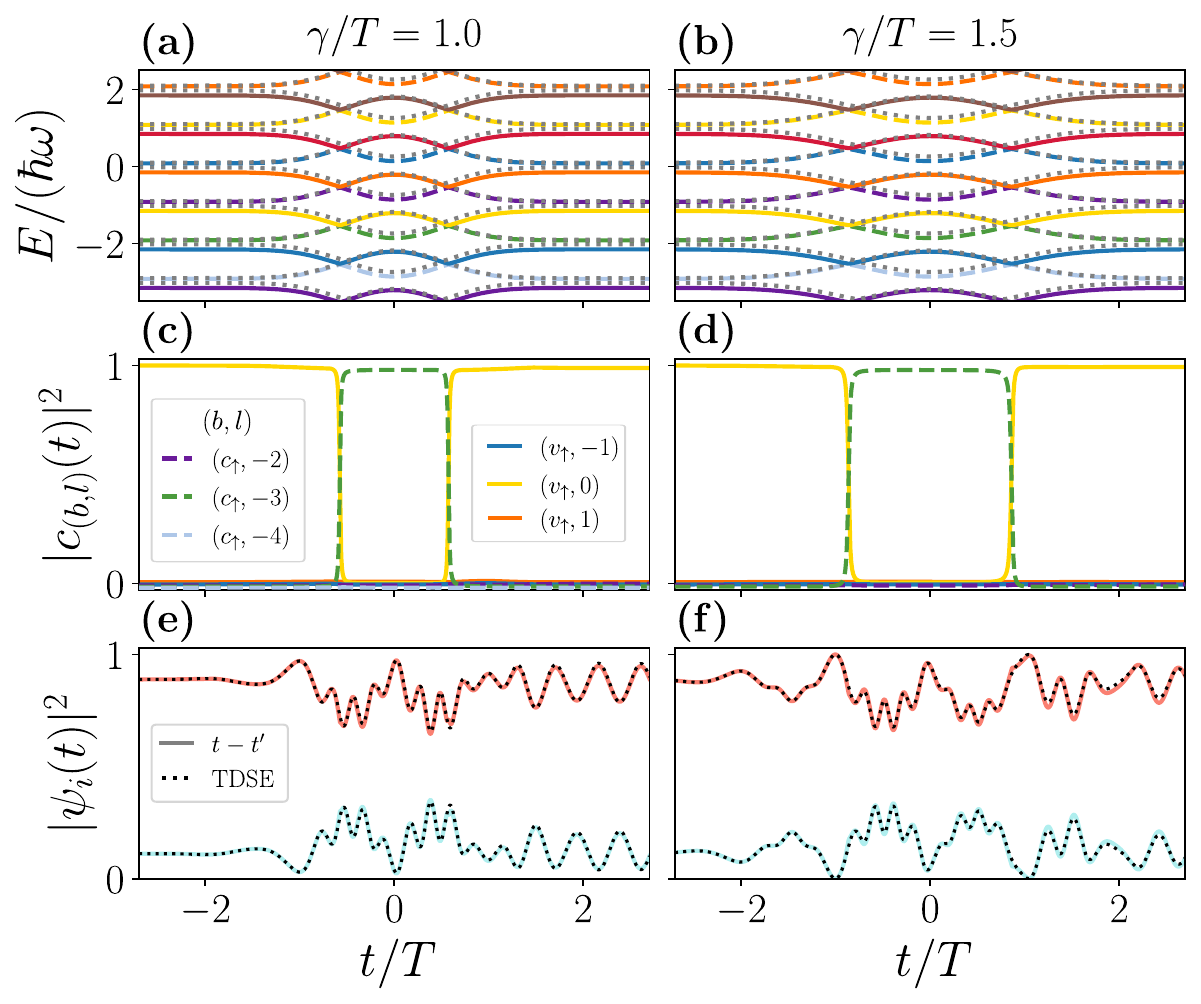}
    \caption{Dependence on the Gaussian width $\gamma$ for $\tau=+1$ and left-handed polarization of a single $(k_x,k_y)=(0.2,0)$ mode, initialized in the state $\ket{v_\uparrow^0}$. The maximum amplitude is fixed to $A_0/(\hbar \omega) = 0.5 $}
    \label{fig:gammaC}
\end{figure}

We include here some complementary figures for the main text. Figure~\ref{fig:gammaC} shows a similar calculation of that presented in figure~\ref{fig:gamma} but for the opposite polarization at fixed valley $\tau = +1$. As discussed in the main text, the coupling of the left-handed polarization is suppressed in this valley in comparison with the right-handed polarization shown in figure~\ref{fig:gamma}. 

\begin{figure}[htb]
    \centering
    \includegraphics[width=\linewidth]{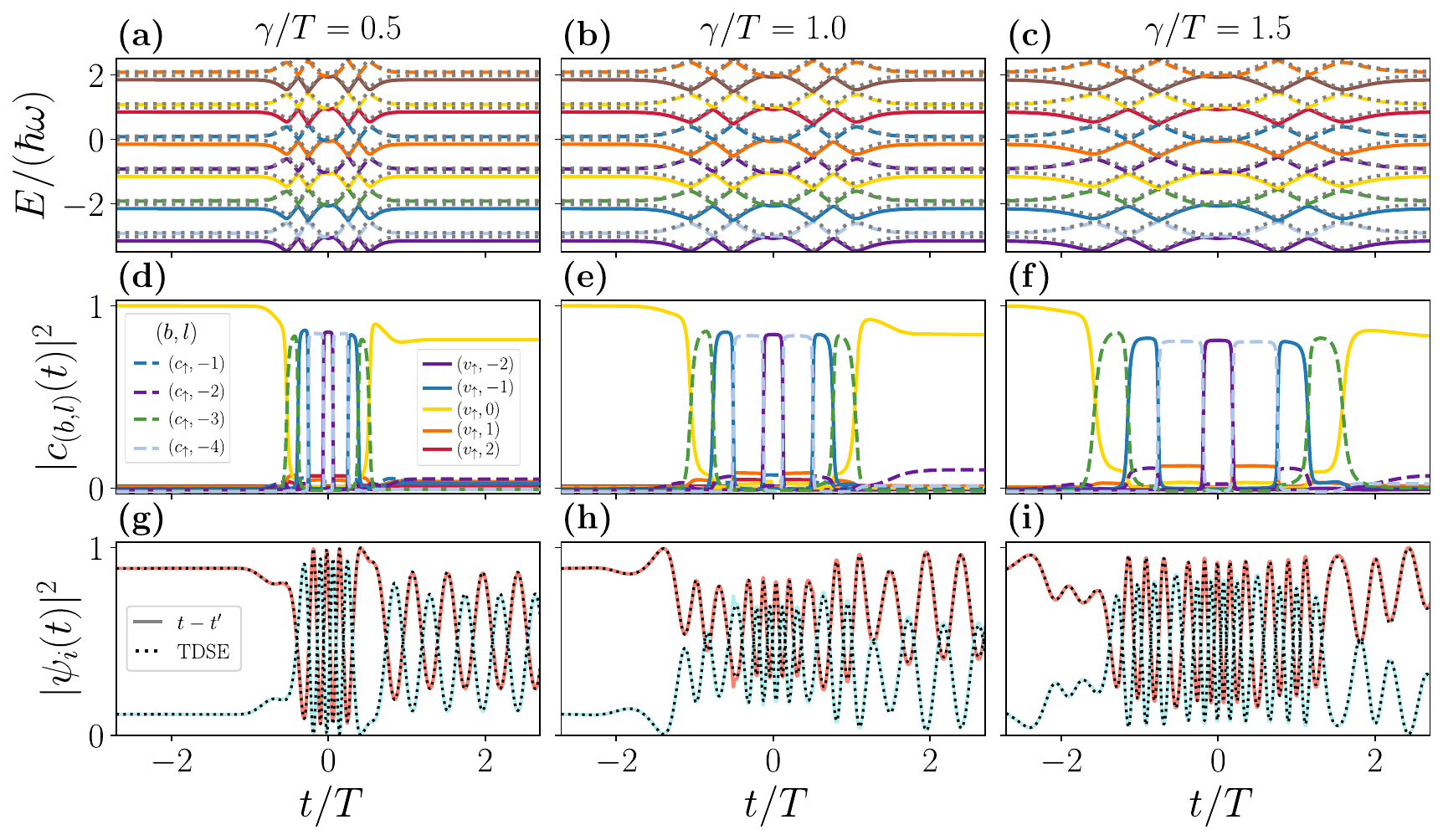}
    \caption{Dependence on the Gaussian width $\gamma$ for $\tau=+1$ and right-handed polarization of a single $(k_x,k_y)=(0.2,0)$ mode, initialized in the state $\ket{v_\uparrow^0}$. The maximum amplitude is fixed to $A_0/(\hbar \omega) = 0.7$.}
    \label{fig:gammaB}
\end{figure}

The results corresponding to a pulse with higher maximum amplitude are shown in figure~\ref{fig:gammaB}. When compared to figure~\ref{fig:gamma}, we note that higher replica are populated. However, the same two regimes in the evolution can be defined: the  adiabatic plateaus and non-adiabatic transitions between replicas, as discussed in the main text regarding figure~\ref{fig:gamma}. 

\begin{figure}[htb]
    \centering
    \includegraphics[width=\linewidth]{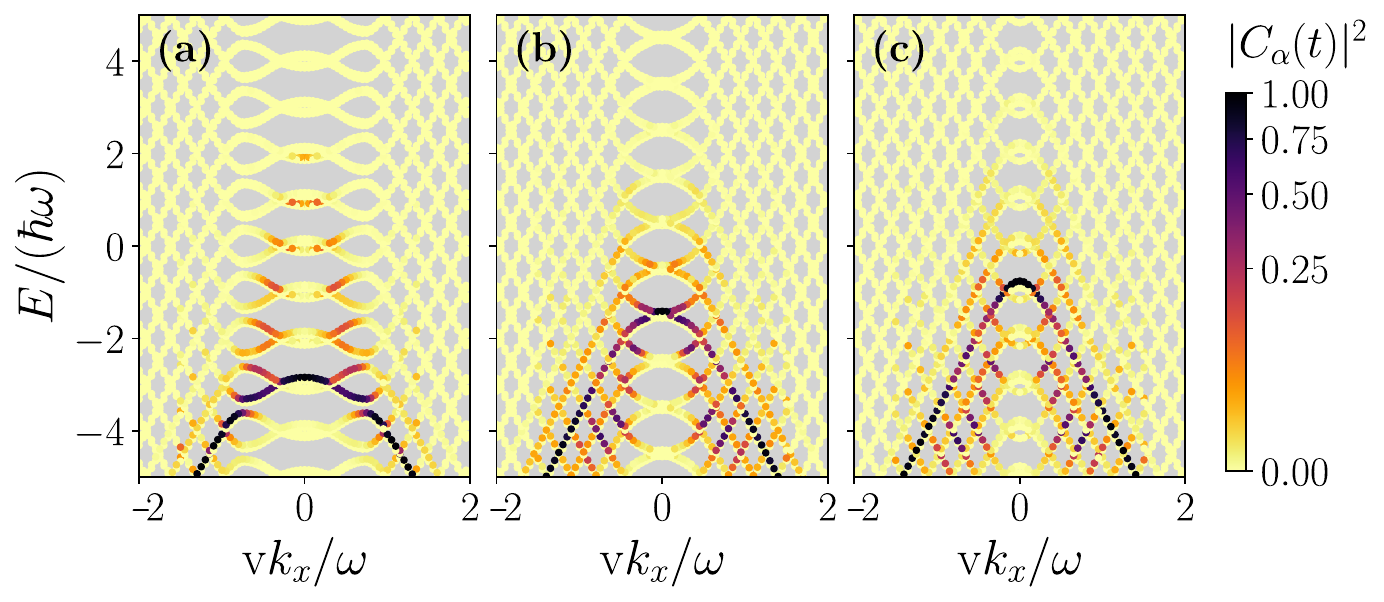}
        \caption{Time-dependent Floquet spectra for the valley $\tau=+1$ under a finite-pulse with right-handed circular polarization ($\theta=0$) and Gaussian envelope centred at $t_0=0$ with a width $\gamma=T$, where $T$ is the pulse period. 
    The spectra are shown for $t/T=\{-0.2,1,2 \}$ in panels (a), (b), (c) and (d), respectively.
    We consider as initial state a fully occupied valence band with spin up $\ket{v_\uparrow^0}$
    and a maximum amplitude $A_0/(\hbar\omega) =0.7$.
    The colour code corresponds to $|C_\alpha(\mathbf{k},t)|^2$. 
    }
    \label{fig:spectrum_b1}
\end{figure}

\begin{figure}[htb]
    \centering
    \includegraphics[width=\linewidth]{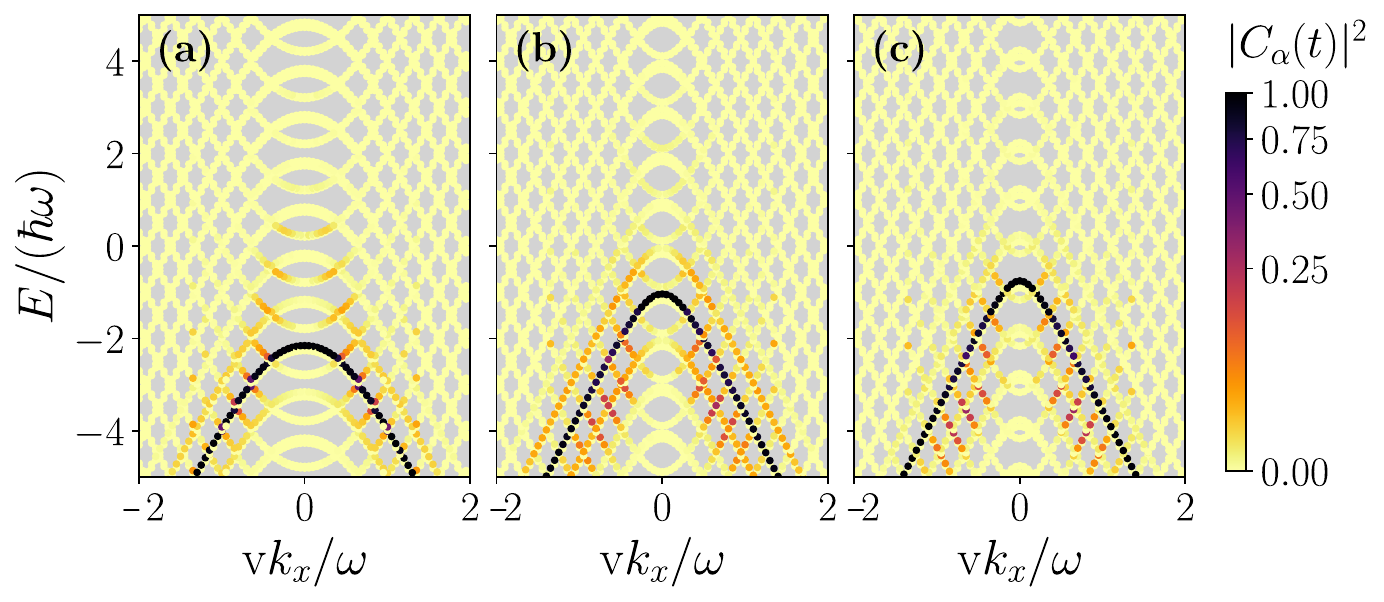}
        \caption{Time-dependent Floquet spectra for the valley $\tau=-1$ under a finite-pulse with right-handed circular polarization ($\theta=0$) and Gaussian envelope centred at $t_0=0$ with a width $\gamma=T$, where $T$ is the pulse period. 
    The spectra are shown for $t/T=\{-0.2,1,2 \}$ in panels~(a), (b), (c) and (d), respectively.
    We consider as initial state a fully occupied valence band with spin down $\ket{v_\downarrow^0}$
 and a maximum amplitude $A_0/(\hbar\omega) =0.7$.
    The colour code corresponds to $|C_\alpha(\mathbf{k},t)|^2$. 
    }
    \label{fig:spectrum_b2}
\end{figure}

Finally, in figures~\ref{fig:spectrum_b1} and \ref{fig:spectrum_b2} complementary time-dependent Floquet spectra are shown for a right-handed polarized pulse and valleys $\tau = \pm 1$. These spectra are calculated for other selected times and complement figures \ref{fig:bandsTTP} and \ref{fig:bandsTTP_tauneg}.

\section*{References}
\providecommand{\newblock}{}


\end{document}